# Chapter **…**
# Strain- and Adsorption-Dependent Electronic States and Transport or Localization in Graphene

**Taras M. Radchenko[1], Igor Yu. Sagalianov[2], Valentyn A. Tatarenko[1],**
**Yuriy I. Prylutskyy[2], Pawel Szroeder[3], Mateusz Kempiński[4,5],**
**and Wojciech Kempiński[6]**

**Abstract**  The chapter generalizes results on influence of uniaxial strain and adsorption on the electron states and charge transport or localization in graphene with different configurations of imperfections (point defects): resonant (neutral) adsorbed atoms either oxygen- or hydrogen-containing molecules or functional groups, vacancies or substitutional atoms, charged impurity atoms or molecules, and distortions. To observe electronic properties of graphene–ad-molecules system, we applied electron paramagnetic resonance technique in a broad temperature range for graphene oxides as a good basis for understanding the electrotransport properties of other active carbons. Applied technique allowed observation of possible metal–insulator transition and sorption pumping effect as well as discussion of results in relation to the granular metal model. The electronic and transport properties are calculated within the framework of the tight-binding model along with the Kubo–Greenwood quantum-mechanical formalism. Depending on electron density and type of the sites, the conductivity for correlated and ordered adsorbates is found to be enhanced in dozens of times as compared to the cases of their random distribution. In case of the uniaxially strained graphene, the presence of point defects counteracts against or contributes to the band-gap opening according to their configurations. The band-gap behaviour is found to be nonmonotonic with strain in case of a simultaneous action of defect ordering and zigzag deformation. The amount of localized charge carriers (spins) is found to be correlated with the content of adsorbed centres (atoms or molecules) responsible for the formation of potential barriers and, in turn, for the localization effects. Physical and chemical states of graphene edges, especially at a uniaxial strain along one of them, play a crucial role in electrical transport phenomena in graphene-based materials.

[1] G. V. Kurdyumov Institute for Metal Physics, N.A.S. of Ukraine, 03142 Kyiv, Ukraine
e-mail: tarad@imp.kiev.ua

[2] Taras Shevchenko National University of Kyiv, 01033 Kyiv, Ukraine

[3] Institute of Physics, Kazimierz Wielki University, 85-090 Bydgoszcz, Poland

[4] Faculty of Physics, Adam Mickiewicz University, 61-614 Poznań, Poland

[5] NanoBioMedical Centre, Adam Mickiewicz University, 61-614 Poznań, Poland

[6] Institute of Molecular Physics, Polish Academy of Sciences, 60-179 Poznań, Poland

## Introduction

Among various types of structural (point or extended) defects in the physics of graphene, adsorbed atoms or molecules are probably the most important examples [1]. They act as the lattice imperfections and strongly affect electronic, optical, thermal, and mechanical properties of graphene. Many characteristics, such as electron states, electrical conductivity, degree of localization of electrons (and their spins), are governed by such defects [2]. Adsorption or introduction of specific defects [3, 4], their configurations (ordering) [5–9], and application of different strains (particularly, uniaxial stretching) [10–13] can serve as ways to solve the problem of gapless graphene for a wide practical application in nanoelectronic devices.

We follow the methodology of the Kubo–Greenwood formalism (see, *e.g.*, reviews [14, 15] and references therein), where transport properties are governed by the movement of electrons. If there are not defects on graphene surface, the electrons can propagate without any backscattering, resembling classical ballistic particles. Therefore, such a transport regime is called the ballistic one. The presence of adsorbed atoms or molecules acting as scattering centres results in diffusive transport regime, when electron diffusion coefficient becomes time-independent, and Ohm's law is valid. Finally, with the course of time, charge carriers start to localize, diffusion coefficient decreases, and localization regime occurs.

Being significantly influenced by adsorption of various atoms and molecules [16, 17], localization process is a crucial issue in the physics of graphene when considering its application in multiple areas such as energy storage, molecule sensing, photovoltaics, and nanoelectronics. This phenomenon can be well observed using the electron paramagnetic resonance (EPR). EPR detects the unpaired spins localized within the structure of material and allows for the observation of their interaction with other spins and the crystal lattice, and was shown as very useful in investigations of graphene-based materials [18, 19].

This chapter summarizes and generalizes the recent theoretical [20, 21] and experimental [22] results obtained for electron behaviours in the afore-mentioned diffusive transport and localization regimes taking place in both unstrained and strained imperfect graphene. Computational results on electron states and quantum transport in diffusive regime are obtained within the framework of both the tight-binding model and the Kubo–Greenwood approach capturing all (ballistic, diffusive, and localization) regimes [14, 15, 23–25]. Experimental observations *via* EPR and conductivity measurements are interpreted using the granular metal model [26] implying appearance of a strong localization of charge carriers due to the existence of potential barriers with charge carrier hopping, which are sensitive to various factors such as temperature, adsorbates, and external fields.

Theoretical part of the study is motivated by, first, disagreements in the literature regarding the stability of differently (randomly, correlatively, or orderly) distributed adatoms of various kinds on graphene surface [27–30], and, second, contradictions concerning impact of the strain on electronic properties of (even perfect, *i.e.*, defect-free) graphene [31–42], all the more so for realistic graphene samples containing different point (or/and extended) defects, particularly, due to the fabrication technology. Experimental part of the work, focusing on graphene oxide and its reduced form, is motivated by, firstly, the current popularity of such materials due to the relative simplicity and repeatability of the manufacturing procedure, and, secondly, their structure imperfection (strong wrinkling, edging, *etc.*), which, being responsible for electronic transport characteristics and localization phenomena, results in more localization sites [22].

## Defect-Configuration-Dependent Charge Carrier Transport

### *Modelling Electronic Transport, Bond Deformations, and Defects*

In the Kubo–Greenwood model, the energy- ($E$) and time- ($t$) dependent diffusivity $D(E,t) = \langle \Delta \hat{X}^2(E,t) \rangle / t$, where the wave-packet mean-quadratic spreading along $x$-direction is $\langle \Delta \hat{X}^2(E,t) \rangle = \text{Tr}[(\hat{X}(t) - \hat{X}(0))^2 \delta(E - \hat{H})] / \text{Tr}[\delta(E - \hat{H})]$ [14, 15] with $\hat{X}(t) = \hat{U}^\dagger(t)\hat{X}\hat{U}(t)$ being the position operator in the Heisenberg representation, $\hat{U}(t) = \exp(-i\hat{H}t/h)$ —time-evolution operator, and tight-binding Hamiltonian (with hopping integrals up to the first 3 coordination shells) define Bernal-stacked few-layer honeycomb lattice [43, 44]. $\hat{H} = \sum_{l=1}^{N_{\text{layer}}} \hat{H}_l + \sum_{l=1}^{N_{\text{layer}}-1} \hat{H}_l'$, where $N_{\text{layer}}$ is a number of layers, $H_l$ is a Hamiltonian contribution of $l$-th layer, and $H_l'$ describes hopping parameters between neighbour layers (vanishes in case of 1 layer) [43, 44], *i.e.*, $\hat{H} = -\gamma_0^1 \sum_{\langle i,j \rangle} c_i^\dagger c_j - \gamma_0^2 \sum_{\langle\langle i,j \rangle\rangle} c_i^\dagger c_j - \gamma_0^3 \sum_{\langle\langle\langle i,j \rangle\rangle\rangle} c_i^\dagger c_j + \sum_i V_i c_i^\dagger c_i$ ; $c_i^\dagger$ and $c_i$ are standard creation or annihilation operators acting on a quasi-particle at the site $i$. The summation over $i$ runs the entire honeycomb lattice, while $j$ is restricted to the nearest-neighbours (in the first term), next nearest-neighbours (second term), and next-to-next nearest-neighbours (third term) of $i$-th site; $\gamma_0^1 = 2.78$ eV is inlayer hopping for the nearest-neighbouring C atoms occupying $i$ and $j$ sites at a lattice-parameter distance $a = 0.142$ nm between them [43, 44]; $\gamma_0^2 = 0.085\gamma_0^1$ and $\gamma_0^3 = 0.034\gamma_0^1$ are intralayer hoppings for the next (second) and next-to-next (third) nearest-neighbouring sites at the second and third coordination shells, respectively [35] (Fig. 1a); and $V_i$ is on-site potential defining defect strength at a given graphene-lattice site $i$ due to the presence of different sources of disorder [43, 44]. The Slonczewski–Weiss–McClure model of electron states [45–47] describes interlayer connection: $\hat{H}_l' = -\gamma_1 \sum_{i,j} (a_{l,j}^\dagger b_{l+1,j} + \text{H.c.}) - \gamma_3 \sum_{j,j'} (b_{l,j}^\dagger a_{l+1,j'} + \text{H.c.})$ with $\gamma_1 = 0.12\gamma_0^1$, $\gamma_3 = 0.1\gamma_0^1$ [44] defining interlayer-hopping amplitudes (Fig. 1a).

We considered two types of uniaxial tensile strain: along so-called armchair

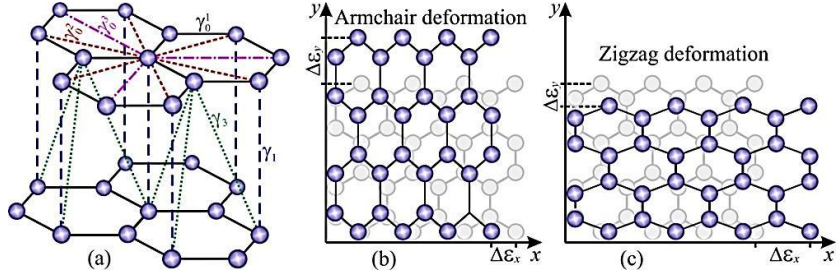

**Fig. 1** Intra- ($\gamma_0^1, \gamma_0^2, \gamma_0^3$) and interlayer ($\gamma_1, \gamma_3$) hopping parameters for two layers (*AB*) of Bernal-stacked multilayer graphene (a). Two types of uniaxial tensile strain (by $\approx 30\%$) along armchair- (b) or zigzag-type (c) edges for single graphene layer.

(Fig. 1b) and zigzag (Fig. 1c) directions (edges). In both cases, the uniaxial strain induces deformation of lattice, *i.e.*, of bond lengths, and hence changes hopping amplitudes between different sites.

Following [35, 36, 48], where random strain is modelled by the Gaussian function, we can obtain dependence of the bond lengths on the deformation tensor components and then relate hopping parameters of the strained ($\gamma$) and unstrained ($\gamma_0^1$) graphene *via* exponential decay: $\gamma(l) = \gamma_0^1 e^{-\beta(l/a-1)}$ with a strained bond length *l*, and decay rate $\beta \approx 3.37$ [35, 36] being extracted from experimental data [49] along with Poisson's ratio $\nu = 0.15$ selected between that measured for graphite [50] and calculated for graphene [51].

We model several kinds of disorder with various point defects. First of them are resonant (uncharged) impurities [43, 44], when C atom from graphene layer is chemically (covalently) bonded with H (C or O) atom from adsorbed organic molecule. Modelling of resonant impurities was carried out with the Hamiltonian part [43, 44] $\hat{H}_{\text{imp}} = \upsilon_d \Sigma_i^{N_{\text{imp}}} d_i^\dagger d_i + V\Sigma_i^{N_{\text{imp}}} (d_i^\dagger c_i + \text{H.c.})$, where $N_{\text{imp}}$ denotes a number of the resonant impurities, and band parameters $V \approx 2\gamma_0^1$ and $\upsilon_d \approx \gamma_0^1/16$ were obtained from density-functional theory calculations [52]. These parameters, being previously adopted for unstrained graphene [43, 44], serve as the typical values for resonant impurities (CH$_3$, C$_2$H$_5$OH, CH$_2$OH as well as hydroxyl groups).

Vacancies are considered as the second important type of defects. A vacancy is regarded as a site with zero hopping parameters to other sites.

Screened charged impurity ions (adatoms, admolecules) on graphene or/and dielectric substrate of it constitute third-type defects. They are commonly described by the Gaussian-type on-site potential [43, 44]: $V_i = \Sigma_{j=1}^{N_{\text{imp}}^V} U_j e^{-|\mathbf{r}_i - \mathbf{r}_j|^2/(2\xi^2)}$, where $\mathbf{r}_i$ is a radius-vector of *i*-th site, $\{\mathbf{r}_j\}$ define positions of $N_{\text{imp}}^V$ impurity atoms (Gaussian centres), $\xi$ is interpreted as an effective potential length, and potential amplitude $U_j$ is uniformly random in the range $[-\Delta, \Delta]$ with $\Delta$—maximum potential magnitude. Varying these parameters, we consider such impurities with shortly ($\xi = 0.65a, \Delta = 3\gamma_0^1$) or more distantly ($\xi = 5a, \Delta = \gamma_0^1$) acting effective potential.

Gaussian hopping [44] is the last type of defects we are interested in. Usually,

they originate from the substitutional impurities causing the atomic-size misfit effect as local in-plane or out-of-plane displacements of atoms, and short- or long-range distortions in graphene lattice due to the curved ripples or wrinkles. In this case, modified distribution of the hopping integrals between different $(i, j)$ sites reads as [44] $\gamma_{i,j} = \gamma + \sum_{k=1}^{N_{hop}^{\gamma}} U_k^{\gamma} e^{-|\mathbf{r}_i - \mathbf{r}_j - 2\mathbf{r}_k|^2/(8\xi_{\gamma}^2)}$ with $N_{hop}^{\gamma}$ (Gaussian) straining centres at the $\mathbf{r}_k$ positions, $\xi_{\gamma}$ is an effective potential length, and hopping amplitude $U_k^{\gamma} \in [-\Delta_{\gamma}, \Delta_{\gamma}]$. The distortion centres are also considered with shortly ($\xi_{\gamma} = 0.65a$, $\Delta\gamma = 1.5\gamma_0^1$) or more distantly acting ($\xi_{\gamma} = 5a$, $\Delta\gamma = 0.5\gamma_0^1$) hoppings. The summation in expressions for Gaussian impurities and hoppings is restricted to the sites belonging to the same layer (possibility for the overlapping of Gaussian distributions in different layers is omitted).

However, sometimes Gaussian (and even Coulomb) scattering potentials are not the most appropriate to describe scattering by various point defects [20]. Therefore, in our calculation, *e.g.*, for K adatoms, we used scattering potential adapted from independent self-consistent *ab initio* calculations [53].

Correlation between impurity adatoms is approximately modelled introducing pair distribution function dependent on correlation length that defines minimal possible distance between any two neighbouring adatoms [8, 20, 23]. Adatomic ordering corresponds to a certain stoichiometric-type superstructure [8, 20].

The dc conductivity $\sigma$ can be extracted from the electron diffusivity $D(E,t)$ undergoing saturation and reaching the maximum, $\lim_{t\to\infty} D(E,t) = D_{max}(E)$, when diffusive transport regime occurs. Then, the semi-classical conductivity at zero temperature is [14, 15, 23] $\sigma = e^2 \tilde{\rho}(E) D_{max}(E)$, where $\tilde{\rho}(E) = \rho/\Omega = \text{Tr}[\delta(E - \hat{H})]/\Omega$ is electron density of sates (DOS) per unit area $\Omega$ (and per spin), and $-e < 0$ denotes the electron charge. The DOS can be used to calculate the electron density as $n_e(E) = \int_{-\infty}^{E} \tilde{\rho}(E) dE - n_{ions}$, where $n_{ions} = 3.9 \cdot 10^{15}$ cm$^{-2}$ is density of positive ions compensating the negative charge of $p$-electrons in graphene. Note that, for the defect-free graphene, at a neutrality (Dirac) point, $n_e(E) = 0$. Combining calculated $n_e(E)$ with $\sigma(E)$, we compute the density-dependent conductivity $\sigma = \sigma(n_e)$. Numerical details for computing DOS, $D$, and $\sigma$ are described in appendices to [23].

## Calculated Density of States, Diffusivities, and Conductivities

Before proceed to graphene with defects of various types, we initially considered defect-free graphene subjected to different values of relative uniaxial tension along above-mentioned two directions. The numerically calculated DOS curves [21] agree with independent analytical results [35]. A spectral gap appearance requires threshold deformations of $\approx 23\%$ along zigzag direction, while there is not any gap opening for any deformations along armchair edge. The band gap opening originates from an additional displacement of both graphene sublattices with respect to each other that occurs most pronouncedly at a deformation along zigzag direction.

High energy values (far from the Dirac point, at $E = 0$) are less experimentally realizable. Therefore, they are not depicted in Fig. 2, where DOS is calculated for single- and bilayer unstrained and strained graphene with 0.1% of random defects.

The DOS curves for graphene monolayer and bilayer (Fig. 2) as well as for tri-layer, quadruple-layer, and quintuple-layer [21] are similar that is an indication of the band-structure similarity, independently of the number of layer. The cause of such similarity lies in the energy band parameters defining intra- and interlayer hopping integrals (see Fig. 1): intralayer nearest-neighbour hoping integral ($\gamma_0^I$) is circa ten times larger than the both interlayer parameters, *i.e.*, interlayer hoppings are much weaker than the intralayer ones.

As Figs 2a and 2d show, resonant impurities (O- or H-containing molecules) and vacancies similarly alter the DOS of the strained graphene: they bring an increase in spectral weight (central peak) near the Dirac point. The central peak, being attributed to impurity (or vacancy) band, increases and broadens as the resonant impurity (or vacancy) concentration rises [21]. The principle distinction between O- or H-containing molecules and vacancies concerning their effects on the spectrum consists in position of the central peak (impurity/vacancy band) in the DOS curves: it is centred at a neutrality point in case of vacancies, whereas it is shifted from it for the hydroxyl groups due to the nonzero (positive) on-site potential modelling them. In contrast to the resonant impurities and vacancies, the Gaussian potentials and hoppings do not induce low-energy impurity (vacancy) band around the neutrality point as shown in Figs 2b, 2c, 2e, 2f.

Like for the perfect graphene [21, 35, 36], the spectrum is also strongly gapless for small and even moderate strains of impure graphene (Fig. 2). The gap over-

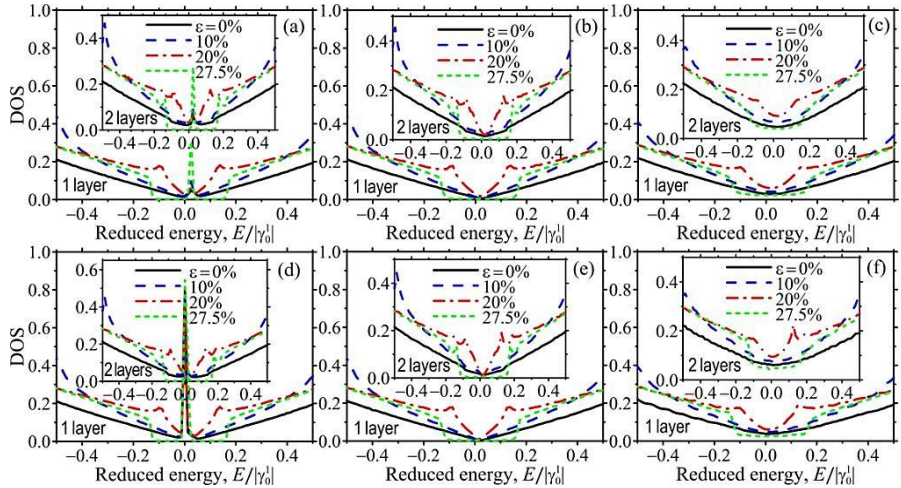

**Fig. 2** DOS for zigzag (un)strained ($0\% \leq \varepsilon \leq 27.5\%$) single- (main panels) and bilayer (insets) graphene with 0.1% of *randomly*-distributed point defects: resonant impurities (a), short- (b) and long-range (c) Gaussian impurities, vacancies (d), short- (e) and long-range (f) Gaussian hoppings.

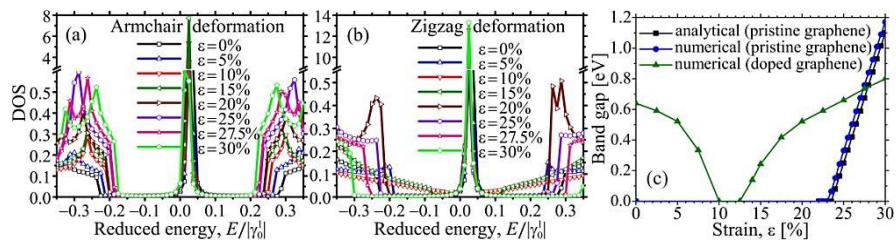

**Fig. 3** (a, b) DOS for graphene monolayer with 3.125% of *ordered* resonant impurities (O- or H-containing molecules) for different (up to 30%) values of the tension strain along armchair (a) and zigzag (b) directions. (c) Comparison of analytically [35] and numerically calculated band gap energies *vs.* the uniaxial deformation along zigzag direction for monolayer graphene without defects (squares and circles) and with 3.125% of ordered hydroxyl groups (triangles).

coming requires the threshold (zigzag) deformations over $\varepsilon \approx 20\%$ for non-long-range acting impurities or vacancies (Figs 2a, 2b, 2d, 2e), whereas 'long-range' potentials (hoppings) smear gap region and transform it into quasi- or pseudo-gap—plateau-shaped deep minimum in DOS near the Dirac point (Figs 2c, 2f).

Figs 3a and 3b show DOS around the Fermi level ($E = 0$) as a function of tensile strain parameter $\varepsilon \in [0\%, 30\%]$ for single-layer graphene with a fixed concentration of the *ordered* H or O adatoms. Band gap decreases slowly (however permanently), if armchair deformation increases. However, in case of zigzag strain, the band gap initially (for $0\% \le \varepsilon \le 10\%$) becomes narrower and narrower up to the total disappearance, but then, at a certain threshold strain value ($\varepsilon_{min} \approx 12.5\%$), the gap reappears, grows up, and can be even wider than it was before the stretching (see also next figure). Importantly, this threshold value $\varepsilon_{min}$, when the band gap opens, is lower in comparison with those that have been estimated earlier for perfect defect-free graphene layers subjected to the uniaxial zigzag strain ($\varepsilon_{min} \approx 23\%$ [35]), shear strain ($\varepsilon_{min} \approx 16\%$ [37]), and almost coincides with the value expected combining shear with armchair uniaxial deformations ($\varepsilon_{min} \approx 12\%$ [37]).

Comparing band-gap energies calculated analytically in [35] and numerically computed for pristine as well as for doped graphene subjected to uniaxial tensile deformation along zigzag-edge direction (Fig. 3c), one can see a pronounced non-monotony of the curve for strained graphene with ordered pattern of defects. Such abnormal nonmonotonic behaviour of the strain-dependent band gap mainly originates from the simultaneous contribution of two factors: impurity ordering and applied strain. Note that numerically obtained curve for defect graphene in Fig. 3c also becomes linear for strains beyond the $\approx 20\%$ and crosses other two curves for pristine graphene close to its predicted failure limit point ($\approx 27.5\%$ [54]).

Due to the honeycomb structure of unstrained graphene lattice, possible adsorption sites can be reduced into three types with high-symmetry favourable (stable) positions; so-called hollow centre (*H*-type), bridge centre (*B*-type), and atop or top (*T*-type) adsorption sites are illustrated in Fig. 4. Taking into account discrepancies in the literature [27–30] on the energy stability (favourableness) of adsorption sites, we study how the positioning of dopants on each *H*, *B*, and *T* site types af-fects the electrotransport properties of unstrained graphene in comparison with the cases of their location on two other types of the sites.

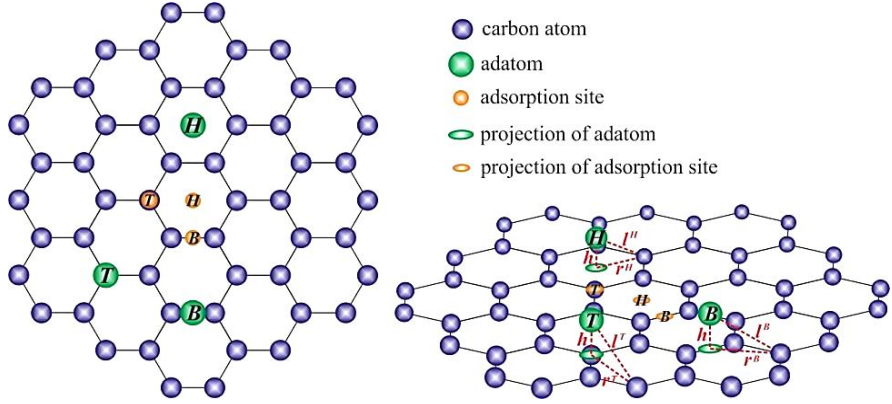

**Fig. 4** Typical adsorption configurations in graphene: top (left) and perspective (right) views of graphene lattice with hollow centre ($H$), bridge centre ($B$), and top ($T$) adsorption sites.

In case of a random adatomic configuration, the steady diffusive regime, when electron diffusivity reaches maximum of it and saturates (Fig. 5a), occurs for a shorter time as compared with correlation (Fig. 5b) and ordering (Fig. 5c) cases. Maximal value in a temporal evolution of diffusivity for ordered impurities substantially exceeds its value for correlated defects and much more for randomly distributed ones. This is a 'hint' that corresponding conductivity should also be higher as compared to other ones. Really, a considerable increase in conductivity due to the correlation and, much more, to the ordering of adatoms as compared with their random distribution is seen from graphs in Figs 6a–6c, where the electron-density dependent conductivities are calculated. The graphs in Figs 6d–6f allow seeing how different types of adsorption sites affect the conductivity for each type of distribution. If adatoms are randomly distributed, conductivity depends on types of adsites: $H$-, $B$-, or $T$-type ones (Fig. 6d). For correlated distribution, conductivity depends on how adatoms manifest themselves: as substitutional (being on $T$-sites) or interstitial (being on $H$- or $B$-sites) atoms (Fig. 6e). If adatoms form ordered superstructures, with equal periods, the conductivity is practically inde-pendent on adsorption type, especially at low electron densities (Fig. 6f).

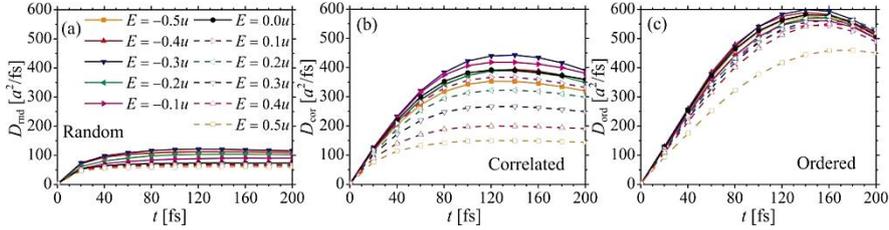

**Fig. 5** Time-dependent diffusivity within the energy range $E \in [-0.5u, 0.5u]$ for random (a), correlated (b), and ordered (c) K adatoms located on hollow ($H$) sites (see also previous figure).

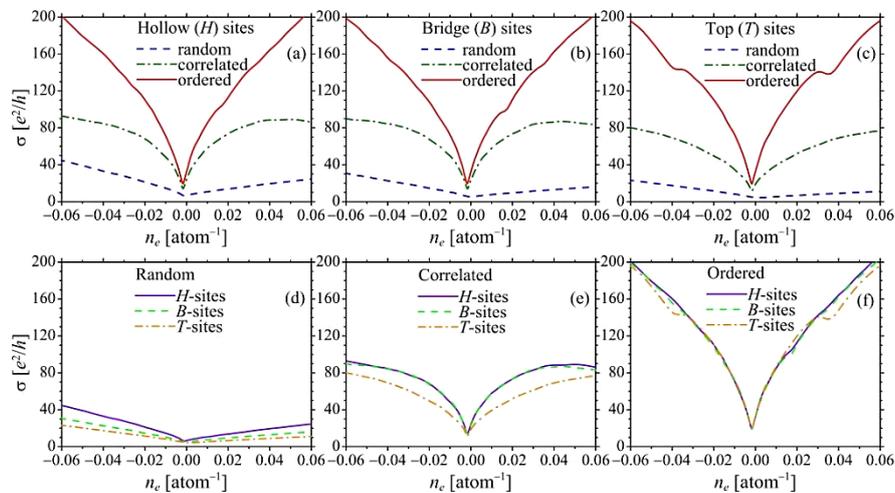

**Fig. 6** Conductivity vs. the electron density for 3.125% of random, correlated, and ordered potassium adatoms occupying hollow (*H*), bridge (*B*), or top (*T*) adsorption sites.

## Adsorption-Driven Charge Carrier (Spin) Localization

### *Sample Preparation and Measurement Conditions*

Graphene oxide (GO) was produced from graphite flakes using the modified Hummers method [55]. Part of this material was consecutively treated with a reducing agent, hydrazine, to obtain the reduced graphene oxide (RGO) [56]. Electron microscopy observation for GO and RGO showed that they are composed of strongly wrinkled microscale sheets as shown in Fig. 7.

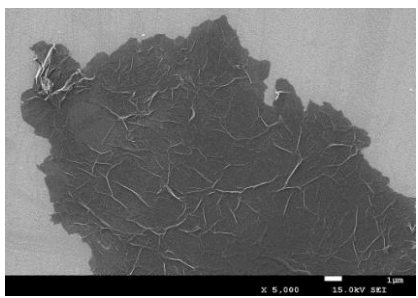

**Fig. 7** Scanning electron microscopy image of a reduced graphene oxide layer (dark area).

Before EPR experiment, both samples were examined with X-ray photoelectron spectroscopy (XPS) for determining the amount of oxygen bound within the structure. XPS experiments showed [22] that the level of functionalization of carbon with oxygen was much higher in GO than RGO.

## Experimental Results and Analysis

Below, we present some features observed in EPR experiments during sequence of the stages: stage 1—purified sample; 2—open to air; 3—purified; 4—open to helium; 5—purified; 6—saturated with heavy water ($D_2O$).

In EPR of RGO experiment, purified RGO showed no EPR signal from carbon, even in the lowest temperatures. There was also no signal from other paramagnetic centres (*e.g.*, Mn ions). Lack of EPR signal of pure RGO in the whole temperature range suggests that electrons are highly delocalized even at the lowest temperatures. The EPR spectra could be observed only after saturation of the sample with guest molecules and decreasing the temperature below 100 K. The comparison of the EPR spectra of RGO after saturation is presented in Fig. 8. Opening the sample tube to air (stage 2) caused the EPR signal of RGO to appear, but only in the low temperature range. Adsorption of guest molecules at the surface of graphene layers hindered the charge carrier transport by creating potential barriers for hopping. Thus, in low temperatures, where the thermal excitations were low, we got localized spins in the system. Introduction of helium resulted in the stronger EPR signal than for the stage 2, most probably because much more helium was adsorbed on the RGO. Saturating the sample with heavy water (stage 6) resulted in the further increase of the EPR signal.

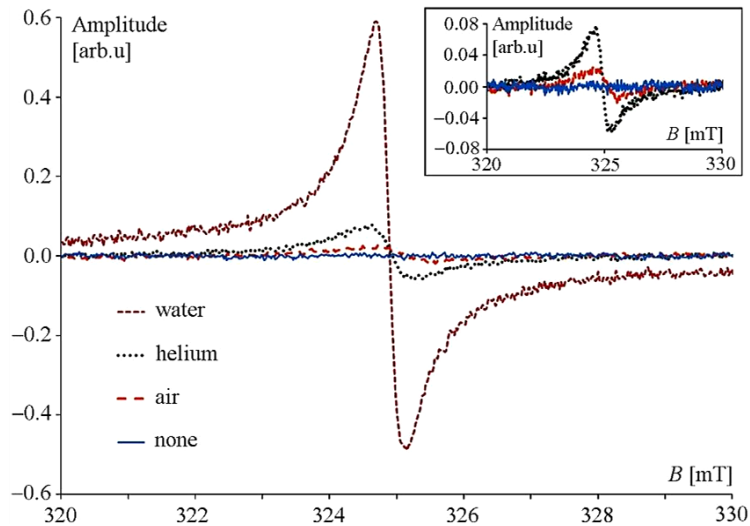

**Fig. 8** EPR spectra of RGO sample in different surroundings at 10 K. The inset shows the magnification of the three low-amplitude signals.

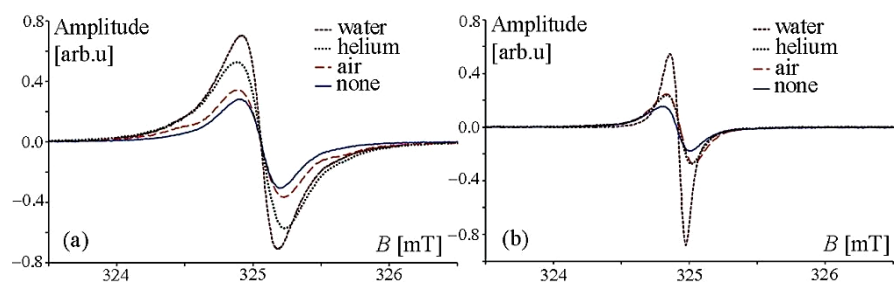

**Fig. 9** EPR spectra of GO sample immersed in various media, recorded at 10 K (a) and room temperature (b).

The EPR spectrum of GO are observed in a whole temperature range at every stage of the sample treatment procedure. Low temperature behaviour, presented in Fig. 9a, was similar to RGO: signal intensity increased according to the sequence: pure–air–helium–water. Striking change appeared in high temperatures (Fig. 9b), where signal amplitudes of the air- and helium-filled sample equated due to the lack of the 'sorption pumping' effect in high temperatures. Above-mentioned observations are clear evidence that electronic properties of graphene-based systems strongly depend on the amount of adsorbed molecules.

The EPR spectra of RGO and GO at 10 K for stages 1 and 6 are shown in Fig.

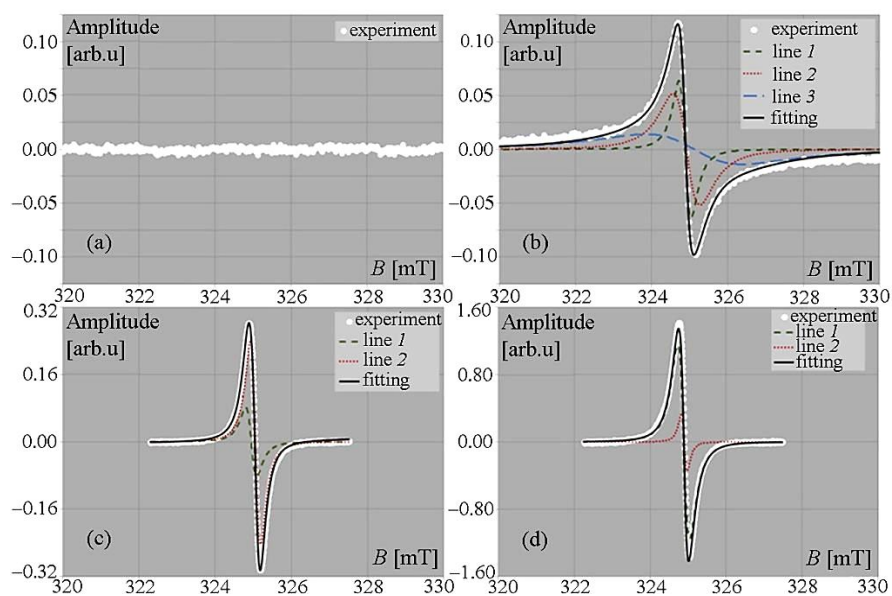

**Fig. 10** EPR spectra of pure RGO (a), RGO + $D_2O$ (b), pure GO (c), and GO saturated with heavy water (d). All spectra were recorded at 10 K.

10. To understand difference between the spectra, note that graphene edges and defects (which are chemically active due to existence of the so-called 'dangling bonds' and show some $sp^3$ hybridization) have significantly different chemical and physical properties than the non-defect layers with $sp^2$ hybridization. Therefore, the paramagnetic centres should also show different behaviour depending on whether they arise from edges (defects) or $sp^2$ planes.

Purified RGO showed no EPR signal at all due to the strong delocalization of charge carriers. However, the signal appeared after adsorption of water, which is possible at the graphene edges with attached functional groups—hydrophilic adsorption sites. Such behaviour could be interpreted as a transition from the 'conducting state' (pure RGO), with lots of percolation paths, to the 'insulating state' (RGO + guest molecules), where thermal excitations were needed to transport the charge across the potential barriers formed at the graphene edges due to the host-guest interactions (hopping transport).

EPR signal of GO occurred in the whole temperature range at both sample treatment stages: purified, and saturated with $D_2O$. The existence of EPR signal resulted from the fact that edges and defects in the graphene layers were terminated with oxygen functional groups and there were no significant areas of well-conducting $sp^2$-graphene.

## Summary and Conclusions

The effects of uniaxial tensile strain and different spatial configurations of adsorptions on electron density of states (DOS) and transport properties of graphene are studied. Spectral gap is sensitive to directions of the tensile strain. The presence of randomly distributed point defects does not avoid the minimum threshold zigzag deformations needed for the band-gap formation. Increase in point-defect concentrations acts against the band-gap opening for all defects considered herein, but their impact is different. However, spatially ordered impurities contribute to the band gap manifestation and can reopen the gap that is normally suppressed by the randomly positioned dopants. Band gap varies nonmonotonically with strain if zigzag deformation and impurity ordering act simultaneously.

For random adatomic distribution on hollow ($H$), bridge ($B$), or top ($T$) sites, the conductivity $\sigma$ depends on their type: $\sigma_{rnd}^H > \sigma_{rnd}^B > \sigma_{rnd}^T$. If adatoms are correlated, $\sigma$ is dependent on whether they act as interstitial or substitutional atoms: $\sigma_{cor}^H \approx \sigma_{cor}^B > \sigma_{cor}^T$. If adatoms form ordered superlattices with equal periods, $\sigma$ is practically independent on the adsorption type: $\sigma_{ord}^H \approx \sigma_{ord}^B \approx \sigma_{ord}^T$. The conductivity for correlated and ordered adatoms is found to be enhanced in dozens of times as compared to the cases of their random positions. The correlation and ordering effects manifest stronger for adatoms acting as substitutional atoms and weaker for those acting as interstitial atoms.

Lack of EPR signal of the purified RGO in the broad temperature range points out that there are not localized spins in the material, even if it is defective ($sp^3$ contribu-tion) and there is some amount of oxygen functional groups attached to the graphene layers. RGO is a good conductor and has no localized spins in pure form. However, adsorption of atoms (molecules) followed by cooling of the system below 100 K resulted in the trapping of charge carriers in the localized states and the appearance of the EPR signal. This behaviour could be interpreted as the adsorption-driven metal–insulator transition. Nevertheless, further research is needed to prove it.

The existence of EPR signal of purified GO is due to the termination of most edges and defects in the graphene layers with oxygen functional groups. The electrical transport was suppressed, making GO an electrical insulator, where localized charge carriers existed even at high temperatures. In this case, adsorption of guest molecules also enhanced localization, with the biggest effect observed for water.

Results for both RGO and GO samples showed that amount of localized charge carriers (spins) correlated with the amount of adsorbed molecules responsible for the formation of potential barriers and, in turn, for the localization effects.

The localization phenomena in graphene-based systems depend heavily on the state of the layer edges, their functionalization and presence of 'foreign' molecules. Both factors can be controlled during and after the material synthesis, which allows for tuning the properties of graphene according to the type of application.

**Acknowledgments** The chapter generalizes results obtained within the framework of Polish–Ukrainian joint research project under the agreement on scientific cooperation between the Polish Academy of Sciences and the National Academy of Sciences of Ukraine for 2015–2017.